\documentclass{aastex62}
\usepackage{rotating}
\usepackage{amsmath}
\usepackage{graphicx}
\newcommand{\hhwang}[1]{{\color{black} #1}}

\received{}
\revised{}
\accepted{}
\submitjournal{ApJ}

\shorttitle{High-Order Accuracy of Self-Gravitational Force Calculation}
\shortauthors{Wang et al.}

\begin{document}

\title{
Self-gravitational Force Calculation of High-Order Accuracy for Infinitesimally Thin Gaseous Disks\footnote{yen@math.fju.edu.tw}}

\correspondingauthor{Chien-Chang Yen}
\email{yen@math.fju.edu.tw}

\author[0000-0002-0786-7307]{Hsiang-Hsu Wang}
\affiliation{
Department of Physics\\
The Chinese University of Hong Kong\\
Shatin, New Territory, Hong Kong, People’s Republic of China}

\author{Ming-Cheng Shiue}
\affiliation{
Department of Applied Mathematics\\
National Chiao Tung University\\
1001 University Road, Hsinchu, Taiwan, R.O.C}

\author{Rui-Zhu Wu}
\affiliation{Department of Mathematics \\
Fu Jen Catholic University\\
New Taipei City, Taiwan}

\author{Chien-Chang Yen}
\affiliation{Department of Mathematics \\
Fu Jen Catholic University\\
New Taipei City, Taiwan}



\begin{abstract}
Self-gravitational force calculation for infinitesimally thin disks is important for studies on the evolution of galactic and protoplanetary disks. Although high-order methods have been developed for hydrodynamic and magneto-hydrodynamic equations, high-order improvement is \hhwang{desirable} for solving self-gravitational forces for thin disks. In this work, we present \hhwang{a new numerical algorithm} that is \hhwang{of linear complexity} and of high-order accuracy. \hhwang{This approach is fast since} the force calculation is \hhwang{associated with} a convolution form, and the fast calculation can be achieved \hhwang{using} Fast Fourier Transform. \hhwang{The nice properties, such as the finite supports and smoothness, of $B$-splines are exploited to \hhwang{stably} interpolate a surface density and achieve a high-order accuracy in forces. }  Moreover, if the mass distribution of interest is \hhwang{exclusively} confined within \hhwang{a} calculation domain,  the method does not require artificial boundary values to be specified before the force calculation.  \hhwang{To validate the proposed algorithm, a series of numerical tests, ranging from 1st- to 3rd-order implementations, are performed and the results are compared with analytic expressions derived for 3rd- and 4th-order \hhwang{generalized} Maclaurin disks.} We conclude that the improvement on the numerical accuracy is significant with the order of the method, with only little increase of the complexity of the method.
\end{abstract}



\section{Introduction} \label{sec:intro}
Evolution of thin disks is of great interest in the context of galactic dynamics and protoplanetary disks formed around young stars. 
The formation of massive stars, the substructures, such as spurs and feathers, associated with barred and spirals galaxies suggest that the self-gravity of gaseous disks plays a role in the evolution of galactic disks \citep{kim12,lee12,lin13,lee14,elm14,seo14}. 
Furthermore, the existence of starburst rings located at the galactic centers also indicates that the rings themselves are gravitationally unstable \citep{hsi11,lin13,seo14,kru15,kru17,seo19}. 
The self-gravity may be also important to planet formation in protoplanetary disks and was suggested to affect the planet radial migration \citep{inu10,zha08,zha14}. 

Although high-order numerical schemes, such as Essentially Non-Oscillatory (ENO; \citet{har87}), Weighted ENO (WENO; \citet{liu94,jia96} ) and Piecewise Parabolic Method (PPM; \citet{col84}),  have been developed for hyperbolic systems (for a recent comprehensive review, see \citet{bal17}), a fast and high-order scheme for self-gravitational force calculation for thin disks is relatively less explored. A direct approach of 2nd-order of accuracy for self-gravitational calculation for infinitesimally thin gaseous disks has been proposed by \citet{yen12}. Within the framework, the approach has been extended to polar coordinates by \citet{wang15}, and adapted to mesh refinement in \citet{wang16}. For gravitational forces in $x$-direction, $F^{x}$,:
\begin{eqnarray}
\label{xforce}
F^{x}(x,y,z)= -\frac{\partial \Phi}{\partial x}(x,y,z)=\int\!\!\!\int\!\!\!\int \frac{\partial {\cal K}}{\partial x}(x-x',y-y',z-z')\rho(x',y',z')\,dx'\,dy'\,dz',
\end{eqnarray}
where $\displaystyle {\cal K}=\frac{1}{\sqrt{x^{2}+y^{2}+z^{2}}}$  is the three-dimensional kernel and the potential satisfies the Poisson equation:
\begin{eqnarray*}
\nabla \cdot (\nabla \Phi)(x,y,z)=4\pi G \rho (x,y,z),
\end{eqnarray*}
where $G$ is the gravitational constant and $\rho$ is the volume density. Relations similar to Eq.~(\ref{xforce}) for $F^x$ can be found for $F^y$ and $F^z$. 
The ``direct" approach has two meanings: one is to calculate the gravitational forces directly, 
instead of taking negative gradient from a pre-calculated gravitational potential, $\Phi(x,y,z)$. 
The other meaning is to compute the integral in Eq.~(\ref{xforce}) directly and reach high-order of accuracy. 

The surface density, $\sigma(x,y)$, of an infinitesimally thin disk relates to the volume density through:
\begin{eqnarray}
\label{surfaceD}
\rho(x,y,z)=\sigma(x,y)\delta(z), 
\end{eqnarray}
where $\delta$ represents the Dirac delta function. \hhwang{Substituting Eq.~(\ref{surfaceD}) into Eq.~(\ref{xforce}) \hhwang{ formally}: 
\begin{eqnarray}
F^{x}(x,y,z)=\int\!\!\!\int\!\!\!\int \frac{\partial {\cal K}}{\partial x}(x-x',y-y',z-z')\sigma(x',y')\delta(z')\,dx'\,dy'\,dz', 
\end{eqnarray}
and integrating over $z'$, the expression of $F^x$ in the plane $z=0$ reads: 
}
\begin{eqnarray}
\label{Fx_convolution}
F^x(x,y,0) =\int\!\!\!\int \frac{\partial {\cal K}}{\partial x}(x-x',y-y',0)\sigma(x',y')\,dx'\,dy'.
\end{eqnarray}
\hhwang{It is important to realize that the singularity involved in the force evaluation, Eq.~(\ref{Fx_convolution}), is removable and integrable.} 
\hhwang{Moreover, Eq.~(\ref{Fx_convolution}) can be theoretically obtained by applying the \hhwang{Lebesgue's dominated} convergence theorem \hhwang{known in measure theory. That is, an infinitesimally thin disk can be approached by a sequence of models with diminishing disk thickness.}} 
In \citet{yen12}, the 2nd-order accuracy is achieved by approximating the a smooth density distribution using Taylor expansion to the linear term in a local sense. The convolution form in Eq.~(\ref{Fx_convolution}) enables the use of Fast Fourier Transform (FFT) that keeps the numerical complexity linear, i.e., $O(N^2\hhwang{\ln^{2}} N)$, where $N$ is the number of zones in one direction. 

The objective of this work is to utilize the spline functions to improve the order of accuracy for force calculations in a general way. This approach involves two parts. The first is to approximate the surface density to high-order accuracy using a set of linearly independent splines, \{$\varphi_{i,j}$\}, where the subscripts $i, j$ are used to distinguish between spline functions. \hhwang{The use of splines for interpolation is well developed in the literature, a rigorous proof and in-depth discussion is beyond the scope of this work. Interested readers are referred to textbooks on numerical analysis \citep{pre75, deB78, epp13}. } The second part involves the calculation in terms of the discretized convolution form. For the first part, the surface density can be approximated by:
\begin{eqnarray}
\label{eqnSigma}
\sigma(x,y)\simeq \sum_{i,j} C_{i,j}\varphi_{i,j}(x,y),
\end{eqnarray}
where $C_{i,j}$ denotes the coefficient associated with the spline $\varphi_{i,j}(x,y)$. Combining Eqs.~(\ref{Fx_convolution}) and (\ref{eqnSigma}), we have:
\begin{eqnarray}
\label{eqnA}
F^{x}(x,y,0)\simeq \sum_{i',j'} C_{i',j'}
\int\!\!\!\int \frac{\partial}{\partial x}{\cal K}(x-x',y-y',0)\varphi_{i',j'}(x',y')dx'\>dy'.
\end{eqnarray}
For fast calculation in the second part, the locations, $(x_i,y_j)$, for evaluating forces are chosen such that Eq.~(\ref{eqnA}) can be recast in a convolution form:
\begin{eqnarray}
\label{eqnF}
F^x_{i,j}= F^x(x_i,y_j,0)\simeq \sum_{i',j'} C_{i',j'} w_{i-i', j-j'},
\end{eqnarray}
where
\begin{eqnarray}
\label{eqnw}
w_{i-i',j-j'}=\int\!\!\!\int \frac{\partial}{\partial x}{\cal K}(x_i-x',y_j-y',0)\varphi_{i',j'}(x',y')dx'\>dy'.
\end{eqnarray}

The paper is organized as follows. In \S~2, we detail the calculation of Eq.~(\ref{eqnw}) for basis spline functions ($B$-splines, hereafter). Generalization to high-order accuracy is presented. \hhwang{In \S~3, we implement the proposed algorithm up to 3rd-order of accuracy and conduct a series of numerical tests. The numerical results are compared with analytic solutions of generalized Maclaurin disks \citep{Sch09,wang15}. 
Finally, we summarize the works and discuss some theoretical issues in \S~4.} 

\section{A direct method using $B$-splines}
\begin{figure}
\epsscale{0.95}
\plotone{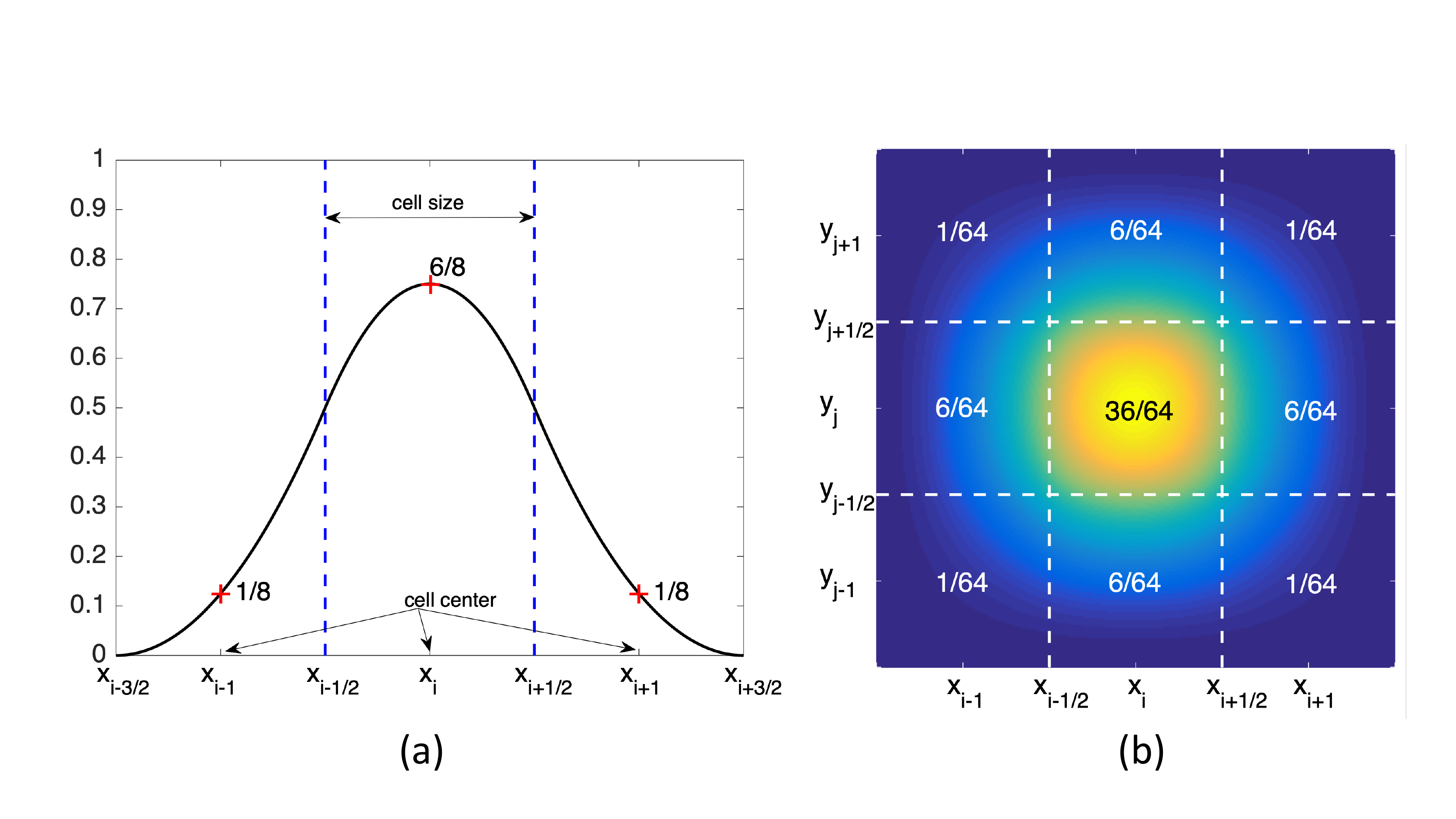}
\caption{\hhwang{(a) One-dimensional profile of a normalized $B$-spline ($k=3$) spanned over a domain $[x_{i-3/2}, x_{i+3/2}]$. The positions of cell boundary are denoted by an half-integer in the subscripts, i.e., $x_{i-3/2}$, $x_{i-1/2}$, $x_{i+1/2}$ and $x_{i+3/2}$. We note that the cell boundaries correspond to the positions of knots $t_i$, $t_{i+1}$, $t_{i+2}$ and $t_{i+3}$ defined in Eq.~(\ref{Bspline3}), respectively. The cell size is defined as the distance between two adjacent cell boundaries, which are shown by the two vertical dashed-lines. The numerical values indicate the weight distribution of a normalized $B$-spline evaluated at cell centers $x_{i-1}$, $x_{i}$  and $x_{i+1}$. According to Eq.~(\ref{Bspline3}), values beyond this domain are zero and not shown in this figure. The second-derivative of this curve exists and is continuous across and beyond the domain. (b) Two-dimensional profile of a normalized $B$-spline tensor spanned over a domain $[x_{i-3/2}, x_{i+3/2}]\times [y_{j-3/2}, y_{j+3/2}]$. This figure is obtained by the tensor product of (a) and the values are color coded. The white dashed-lines represent the cell boundaries. The values indicate the weight distribution evaluated at cell centers and collectively form the weight matrix Eq~(\ref{Eqn:weight_matrix}).  We note that values beyond this domain are zero, which is a direct consequence of tensor product from (a). }}
\label{fig:Bspline_figure}
\end{figure}

In this work, we adopt the $B$-splines as the basis functions to interpolate surface density. The first order of $B$-splines in one dimension reads:
\begin{eqnarray}
\label{Bspline1}
B^{[1]}_i(x)\equiv\left\{
\begin{array}{cc}
1 ,& \mbox{ if } t_i \le x < t_{i+1}\\
0 ,& \mbox{ otherwise}
\end{array}
\right. ,
\end{eqnarray}
where the upper script represents the order of $B$-splines and the set $\{t_i\}$ (a.k.a knots) denotes the locations where different $B$-splines meet. \hhwang{Conventionally, $\{t_i\}$ is sorted in a non-decreasing order.} The $B$-spline of order $k>1$ is defined recursively:
\begin{eqnarray}
\label{eqnBk}
B^{[k]}_i(x)\equiv
\frac{x-t_i}{t_{i+k-1}-t_i}B^{[k-1]}_i(x)
+\frac{t_{i+k}-x}{t_{i+k}-t_{i+1}}B^{[k-1]}_{i+1}(x).
\end{eqnarray}
Based on the definition of $B$-splines, the support of $B^{[k]}_i(x)$ is $[t_i,t_{i+k})$. For $k=2$ and $k=3$, the explicit expressions are:
\begin{eqnarray}
\label{Bspline2}
B^{[2]}_i (x)=
\left\{
\begin{array}{ll}
\displaystyle\frac{(x-t_i)}{(t_{i+1}-t_i)}, & \mbox{ if } t_i\le x < t_{i+1}\\
\displaystyle\frac{(t_{i+2}-x)}{(t_{i+2}-t_{i+1})},
& \mbox{ if } t_{i+1}\le x< t_{i+2} \\
\displaystyle 0,
& \mbox{otherwise}
\end{array}
\right. ,
\end{eqnarray}
and
\begin{eqnarray}
\label{Bspline3}
B^{[3]}_i (x)=
\left\{
\begin{array}{ll}
\displaystyle\frac{(x-t_i)^2}{(t_{i+1}-t_i)(t_{i+2}-t_i)}, & \mbox{ if } t_i\le x < t_{i+1}\\

\displaystyle\frac{(x-t_i)(t_{i+2}-x)}{(t_{i+2}-t_i)(t_{i+2}-t_{i+1})}
+ \frac{(x-t_{i+1})(t_{i+3}-x)}{(t_{i+2}-t_{i+1})(t_{i+3}-t_{i+1})},
&\mbox{ if } t_{i+1}\le x < t_{i+2}\\
\displaystyle\frac{(t_{i+3}-x)^2}{(t_{i+3}-t_{i+1})(t_{i+3}-t_{i+2})},
& \mbox{ if } t_{i+2}\le x< t_{i+3} \\
\displaystyle 0,
& \mbox{otherwise}
\end{array}
\right. ,
\end{eqnarray}
respectively. In this work, the values of $t_i$ are all distinct, thus the first $k-1$ derivatives of $B$-splines are continuous across $t_i$. It is the finite support and the smoothness \hhwang{properties} that make $B$-splines a good choice for approximating the surface density in Eq.~(\ref{eqnSigma}). The two-dimensional basis set is formed using the tensor product $\varphi_{i,j}(x,y)=B^{[k]}_i(x)B^{[k]}_j(y)$. Substituting this expression of $\varphi_{i,j}(x,y)$ into Eq.~(\ref{eqnw}), it follows:
\begin{eqnarray}
\label{eqnwB}
w_{i-i',j-j'}=\int\!\!\!\int \frac{\partial}{\partial x}{\cal K}(x-x',y-y',0)B^{[k]}_{i'}(x')B^{[k]}_{j'}(y')dx'\>dy',
\end{eqnarray}
for $(x,y)=(x_i,y_j)$. Suppose that the surface density is confined within a finite rectangular computational domain ${\cal D}=[L,R]\times[B,T]$, where $L<R$ and $B<T$, and the domain is uniformly subdivided by rectangular cells, with $\Delta x$ and $\Delta y$ being the cell sizes in $x$- and $y$-direction, respectively. The grid points are $x_{i+1/2}=L+i\Delta x$ for $i=0,1,...,M$ and $y_{j+1/2}=B+j\Delta y$ for $j=0,1,...,N$. The cell center at $(i,j)$ is $\displaystyle (x_i,y_j)=(\frac{x_{i-1/2}+x_{i+1/2}}{2}, \frac{y_{j-1/2}+y_{j+1/2}}{2})$. \hhwang{Figure~\ref{fig:Bspline_figure}a shows a spline $B^{[3]}(x)$ peaked at the cell center $x_i$, with $x_{i-1/2}$ and $x_{i+1/2}$ being the cell boundaries. For this particular $B$-spline of $k=3$, the size of its support is $3\Delta x$ and its first  and second derivatives are continuous over the whole computational domain. Figure~\ref{fig:Bspline_figure}b shows the tensor product $B^{[3]}(x)B^{[3]}(y)$ peaked at $(x_i, y_j)$, with white dashed-lines indicating the cell boundaries located at half-integer subscripts. The map of this tensor are color-coded and the values shown in the figure are evaluated for the cell center $(x_i, y_j)$ (black-\hhwang{font} number) and for those centers surrounding it (white-\hhwang{font} numbers).}

\subsection{Calculation of Equation~(\ref{eqnwB})}
As will be shown in \S~2.2, the calculation of Eq.~(\ref{eqnwB}) involves an integration of the form:
\hhwang{
\begin{eqnarray}
F(x,y,\alpha,\beta)&=&\int\!\!\!\int \frac{x^\alpha y^\beta}{(x^2+y^2)^{3/2}}\,dx\>dy \\
&=&\frac{y^{\beta+1}}{\alpha+\beta-1}\int \frac{x^{\alpha}}{(x^2+y^2)^{3/2}}\,dx + \frac{x^{\alpha+1}}{\alpha+\beta-1}\int \frac{y^{\beta}}{(x^2+y^2)^{3/2}}\,dy, \label{intF}\\
&=&\frac{y^{\beta+1}}{\alpha+\beta-1}G(x,y,\alpha) + \frac{x^{\alpha+1}}{\alpha+\beta-1}G(y,x,\beta),
\end{eqnarray}
where $\alpha$ and $\beta$ are non-negative integers (see Appendix B for the derivation of Eq.~(\ref{intF})) and 
\begin{eqnarray}
G(x,a,\alpha)=\int \frac{x^{\alpha}}{(x^{2}+a^{2})^{3/2}}\,dx. \label{G_function}
\end{eqnarray}
}
 \hhwang{Equation~(\ref{G_function}) can be evaluated through the following recursion relation}:
\hhwang{
\begin{eqnarray}
\label{intMoment}
G(x,y,\alpha)=\frac{x^{\alpha-1}}{\alpha-2}\frac{1}{(x^2+y^2)^{1/2}}-\frac{\alpha-1}{\alpha-2}y^2 G(x,y,\alpha-2).
\end{eqnarray}
}
The derivation of Eq.~(\ref{intMoment}) is given in Appendix B.
Although Eq.~(\ref{intF}) appears to apply only when $\alpha+\beta > 1$, this is not a restrictive limitation since the close forms for $\alpha+\beta=0,1,2,3,4$ exist and are listed in Appendix A\footnote{Symbolic mathematical software, {Maple}, fails to find the close form for $(\alpha,\beta)=(4,0), (2,2)$ and $(0,4)$}. For the convenience of discussion below, we define $\tilde{F}(x_{\ell},x_{r},y_{\ell},y_{r},\alpha, \beta)$ to be the definite integral of $F(x,y,\alpha,\beta)$ over the domain $[x_{\ell},x_{r}]\times [y_{\ell},y_{r}]$:
\begin{eqnarray}
\label{intFtilde}
\tilde{F}(x_{\ell},x_{r},y_{\ell},y_{r},\alpha, \beta)&=&\int_{x_{\ell}}^{x_{r}}\int_{y_{\ell}}^{y_r} \frac{x^\alpha y^\beta}{(x^2+y^2)^{3/2}}\,dy\>dx.
\end{eqnarray}

\subsection{Examples}

\noindent {\sl {\bf Case 1.}} ($k=1$) 

For $(x,y)= (x_i,y_j)$ and from Eq.~(\ref{Bspline1}), Eq.~(\ref{eqnwB}) reads:
\begin{eqnarray}
w_{i-i',j-j'}&=&\int_{x_{i'-1/2}}^{x_{i'+1/2}}\int_{y_{j'-1/2}}^{y_{j'+1/2}} \frac{\partial}{\partial x}{\cal K}(x-x',y-y',0)dy'\>dx' \nonumber \\
&=& \int_{x_{i'-1/2}}^{x_{i'+1/2}}\int_{y_{j'-1/2}}^{y_{j'+1/2}} \frac{x'-x}{[(x-x')^2+(y-y')^2]^{3/2}}dy'\>dx' \nonumber \\
&=& \int_{x_{i',0}}^{x_{i',1}}\int_{y_{j',0}}^{y_{j',1}} \frac{\tilde{x}}{(\tilde{x}^2+\tilde{y}^2)^{3/2}}d\tilde{y}\>d\tilde{x}, \label{wk1}
\end{eqnarray}
where in Eq.~(\ref{wk1}) we have applied the change of variable $(\tilde{x},\tilde{y})=(x'-x, y'-y)$ and introduced auxiliary variables $x_{i',r} = x_{i'-1/2+r}-x$ and $y_{j',r} = y_{j'-1/2+r}-y$ for $r=0,..,k$ to denote the ranges of integration. Eq.~(\ref{wk1}) is simply $\tilde{F}(x_{i',0},x_{i',1},y_{j',0},y_{j',1},1, 0)$. In Table~\ref{tbl_Fk1} the integration Eq.~(\ref{wk1}) is presented in a more concise way. Each entry in the table denotes a triplet $(\alpha,\beta, \gamma)$, where $\gamma$ is the associated coefficient. 

\hhwang{In general, as will be shown and become more clear in the following cases, if one attempts to expand the integral $w_{i-i',j-j'}$ term by term, each term would end up with a form:
\begin{eqnarray}
\gamma \tilde{F}(x_{\ell},x_{r},y_{\ell},y_{r},\alpha, \beta)&=&\gamma \int_{x_{\ell}}^{x_{r}}\int_{y_{\ell}}^{y_r} \frac{x^\alpha y^\beta}{(x^2+y^2)^{3/2}}\,dy\>dx.
\end{eqnarray}
Each term has a different range of integration (see the definition of $B$-splines but generalize it for a tensor in 2D), different $(\alpha, \beta)$ and a constant coefficient $\gamma$ associated with it. Take $k=3$ for example,  one would expect nine different integration ranges (three in $x$ and three in $y$, respectively; see also the nine blocks shown in Fig.~\ref{fig:Bspline_figure}b), which corresponds the size of Table~\ref{tbl_Fk3}. Within one integration range, one would expect nine terms associated with different $(\alpha, \beta)$ and a coefficient $\gamma$ attached to each of them. In the following cases, to avoid the clutter but straightforward formulas, we summarize that information in a table for each case. 
}

\noindent {\sl {\bf Case 2.}} ($k=2$)

For $(x,y)=(x_i,y_j)$ and from Eq.~(\ref{Bspline2}), Eq.~(\ref{eqnwB}) reads:
\begin{eqnarray}
w_{i-i',j-j'}&=&\int_{x_{i'-1/2}}^{x_{i'+3/2}}\int_{y_{j'-1/2}}^{y_{j'+3/2}} \frac{\partial}{\partial x}{\cal K}(x-x',y-y',0)B^{[2]}_{i'}(x')B^{[2]}_{j'}(y')dy'\>dx' \nonumber \\
&=& \int_{x_{i',0}}^{x_{i',1}}\int_{y_{j',0}}^{y_{j',1}} \frac{\tilde{x}}{(\tilde{x}^2+\tilde{y}^2)^{3/2}}\frac{(\tilde{x}-x_{i',0})(\tilde{y}-y_{j',0})}{\Delta x \Delta y}d\tilde{y}\>d\tilde{x} \nonumber \\
&+& \int_{x_{i',1}}^{x_{i',2}}\int_{y_{j',0}}^{y_{j',1}} \frac{\tilde{x}}{(\tilde{x}^2+\tilde{y}^2)^{3/2}}\frac{(x_{i',2}-\tilde{x})(\tilde{y}-y_{j',0})}{\Delta x \Delta y}d\tilde{y}\>d\tilde{x} \nonumber \\
&+& \int_{x_{i',0}}^{x_{i',1}}\int_{y_{j',1}}^{y_{j',2}} \frac{\tilde{x}}{(\tilde{x}^2+\tilde{y}^2)^{3/2}}\frac{(\tilde{x}-x_{i',0})(y_{j',2}-\tilde{y})}{\Delta x \Delta y}d\tilde{y}\>d\tilde{x} \nonumber \\
&+& \int_{x_{i',1}}^{x_{i',2}}\int_{y_{j',1}}^{y_{j',2}} \frac{\tilde{x}}{(\tilde{x}^2+\tilde{y}^2)^{3/2}}\frac{(x_{i',2}-\tilde{x})(y_{j',2}-\tilde{y})}{\Delta x \Delta y}d\tilde{y}\>d\tilde{x}. \label{wk2}
\end{eqnarray}
The associated integrations are listed in Table~\ref{tbl_Fk2}. We note that the final result would need to sum over all the integration results listed in Table~\ref{tbl_Fk2} and multiply by a factor $1/(\Delta x \Delta y)$. 

\noindent {\sl {\bf Case 3.}} ($k=3$)
With the recursion relation Eq.~(\ref{intF}), it is straightforward to generalize to a third-order ($k=3$) of accuracy or higher. For $(x,y)=(x_i,y_j)$ and from Eq.~(\ref{Bspline3}), Eq.~(\ref{eqnwB}) reads:
\begin{eqnarray}
w_{i-i',j-j'}&=&\int_{x_{i'-1/2}}^{x_{i'+5/2}}\int_{y_{j'-1/2}}^{y_{j'+5/2}} \frac{\partial}{\partial x}{\cal K}(x-x',y-y',0)B^{[3]}_{i'}(x')B^{[3]}_{j'}(y')dy'\>dx' \nonumber \\
&=& \frac{1}{4(\Delta x)^2(\Delta y)^2}\int_{x_{i',0}}^{x_{i',1}}\int_{y_{j',0}}^{y_{j',1}} \frac{\tilde{x}}{(\tilde{x}^2+\tilde{y}^2)^{3/2}} (\tilde{x}-x_{i',0})^2(\tilde{y}-y_{j',0})^2d\tilde{y}\>d\tilde{x} \nonumber \\
&+& \frac{1}{4(\Delta x)^2(\Delta y)^2}\int_{x_{i',1}}^{x_{i',2}}\int_{y_{j',0}}^{y_{j',1}} \frac{\tilde{x}}{(\tilde{x}^2+\tilde{y}^2)^{3/2}} [(\tilde{x}-x_{i',0})(x_{i',2}-\tilde{x})+(\tilde{x}-x_{i',1})(x_{i',3}-\tilde{x})](\tilde{y}-y_{j',0})^2d\tilde{y}\>d\tilde{x} \nonumber \\ 
&+& \frac{1}{4(\Delta x)^2(\Delta y)^2}\int_{x_{i',2}}^{x_{i',3}}\int_{y_{j',0}}^{y_{j',1}} \frac{\tilde{x}}{(\tilde{x}^2+\tilde{y}^2)^{3/2}} (x_{i',3}-\tilde{x})^2(\tilde{y}-y_{j',0})^2d\tilde{y}\>d\tilde{x} \nonumber \\ 
&+& \frac{1}{4(\Delta x)^2(\Delta y)^2}\int_{x_{i',0}}^{x_{i',1}}\int_{y_{j',1}}^{y_{j',2}} \frac{\tilde{x}}{(\tilde{x}^2+\tilde{y}^2)^{3/2}} (\tilde{x}-x_{i',0})^2[(\tilde{y}-y_{j',0})(y_{j',2}-\tilde{y})+(\tilde{y}-y_{j',1})(y_{j',3}-\tilde{y})]d\tilde{y}\>d\tilde{x} \nonumber \\
&+& \frac{1}{4(\Delta x)^2(\Delta y)^2}\int_{x_{i',1}}^{x_{i',2}}\int_{y_{j',1}}^{y_{j',2}} \frac{\tilde{x}}{(\tilde{x}^2+\tilde{y}^2)^{3/2}}[(\tilde{x}-x_{i',0})(x_{i',2}-\tilde{x})+(\tilde{x}-x_{i',1})(x_{i',3}-\tilde{x})] \nonumber \\ 
&&\times [(\tilde{y}-y_{j',0})(y_{j',2}-\tilde{y})+(\tilde{y}-y_{j',1})(y_{j',3}-\tilde{y})]d\tilde{y}\>d\tilde{x} \nonumber \\ 
&+& \frac{1}{4(\Delta x)^2(\Delta y)^2}\int_{x_{i',2}}^{x_{i',3}}\int_{y_{j',1}}^{y_{j',2}} \frac{\tilde{x}}{(\tilde{x}^2+\tilde{y}^2)^{3/2}} (x_{i',3}-\tilde{x})^2[(\tilde{y}-y_{j',0})(y_{j',2}-\tilde{y})+(\tilde{y}-y_{j',1})(y_{j',3}-\tilde{y})]d\tilde{y}\>d\tilde{x} \nonumber  \\
&+& \frac{1}{4(\Delta x)^2(\Delta y)^2}\int_{x_{i',0}}^{x_{i',1}}\int_{y_{j',2}}^{y_{j',3}} \frac{\tilde{x}}{(\tilde{x}^2+\tilde{y}^2)^{3/2}} (\tilde{x}-x_{i',0})^2(y_{j',3}-\tilde{y})^2d\tilde{y}\>d\tilde{x} \nonumber \\ 
&+& \frac{1}{4(\Delta x)^2(\Delta y)^2}\int_{x_{i',1}}^{x_{i',2}}\int_{y_{j',2}}^{y_{j',3}} \frac{\tilde{x}}{(\tilde{x}^2+\tilde{y}^2)^{3/2}} [(\tilde{x}-x_{i',0})(x_{i',2}-\tilde{x})+(\tilde{x}-x_{i',1})(x_{i',3}-\tilde{x})](y_{j',3}-\tilde{y})^2d\tilde{y}\>d\tilde{x} \nonumber \\
&+& \frac{1}{4(\Delta x)^2(\Delta y)^2}\int_{x_{i',2}}^{x_{i',3}}\int_{y_{j',2}}^{y_{j',3}} \frac{\tilde{x}}{(\tilde{x}^2+\tilde{y}^2)^{3/2}} (x_{i',3}-\tilde{x})^2(y_{j',3}-\tilde{y})^2d\tilde{y}\>d\tilde{x} \label{wk3}
\end{eqnarray}

The calculation for $k=3$ involves 81 integrations as listed in Table~\ref{tbl_Fk3}. We emphasize that, the complicated evaluation of Eq.~(\ref{eqnwB}) need only to be done once at the beginning of simulations, and the results can be reused afterwards when needed. We note that the final result would need to sum over all the integration results listed in Table~\ref{tbl_Fk3} and multiply by a factor $1/(4(\Delta x)^2(\Delta y)^2)$. 

\subsection{Solving a linear system of $C_{i,j}$}
Using the $B$-splines as the basis functions, Eq.~(\ref{eqnSigma}) now reads:
\begin{eqnarray}
\sigma(x,y)\simeq \sum_{i,j} C_{i,j}B^{[k]}_i(x)B^{[k]}_j(y). 
\end{eqnarray}
That is, we are using $B$-splines to interpolate the surface density. By construction, a $B$-spline has a finite support and a symmetric point. Except for $k=1$ (piecewise constant), the symmetric center corresponds to the location of  maximum. We found that it is advantageous to have the symmetric centers of $B$-splines coincide with cell centers, $(x_i, y_j)$. Applying the condition:
\begin{eqnarray}
\label{eqnSigmaB}
\sigma(x_i,y_j) = \sum_{i',j'} C_{i',j'}B^{[k]}_{i'}(x_i)B^{[k]}_{j'}(y_j), 
\end{eqnarray}
we are solving an unknown matrix $C_{i,j}$ with a size the same as  $\sigma_{i,j}\equiv \sigma(x_i,y_j)$.  Due to the finite support of $B$-splines, unless $k$ is large, usually only those $B$-splines in the vicinity of $(x_i, y_j)$ are  involved. For example, when $k=1,2$, one may simply take $C_{i,j}=\sigma(x_i,y_j)$ given that $B^{[k]}_i(x)B^{[k]}_j(y)$ is normalized.  For higher order $B$-splines, we adopt $k=3$ to detail the procedure as follows. We define a weight matrix:
\begin{equation}
W
=[W_{i,j}]_{{1\le i,j\le 3}}=
\frac{1}{64}\begin{bmatrix}
    1 & 6 & 1 \\
    6 & 36 & 6 \\
    1 & 6 & 1 
\end{bmatrix}, 
\label{Eqn:weight_matrix}
\end{equation}
where the factor in front of the matrix is used to normalize the matrix such that the summation over all the entries of W is unity. The entries in $W$ describe the weight distribution of a $B$-spline tensor $B^{[3]}_i(x)B^{[3]}_j(y)$ for the cell centers located within its support. \hhwang{A visual interpretation of the weight matrix is shown in Fig.~\ref{fig:Bspline_figure}b, which describes how a $B^{[3]}(x)B^{[3]}(y)$ tensor peaked at $(x_i, y_j)$ contributes to itself and those cell centers adjacent to it. If a set of $B$-spline tensors is used to interpolate a density located at $(x_i,y_j)$, it is obvious that only those nine tensors surrounding $(x_i, y_j)$ will contribute. The finite support of $B$-spline therefore greatly simplifies the linear system for density interpolation. For a linear system of $k=3$, the surface density $\sigma_{i,j}$ has the following relation with $C_{i,j}$: }
\hhwang{
\begin{eqnarray}
\label{Eqn:sigmaW}
\sigma_{i,j}&=&C_{i-1,j+1}W_{33}+C_{i,j+1}W_{32}+C_{i+1,j+1}W_{31} \nonumber \\
                  &+&C_{i-1,j}W_{23}+C_{i,j}W_{22}+ C_{i+1,j}W_{21} \nonumber \\
                  &+&C_{i-1,j-1}W_{13}+C_{i,j-1}W_{12}+C_{i+1,j-1}W_{11} \nonumber \\
                  &=& \frac{1}{64}C_{i-1,j+1}+\frac{6}{64}C_{i,j+1}+\frac{1}{64}C_{i+1,j+1} \nonumber \\
                  &+& 6(\frac{1}{64}C_{i-1,j}+\frac{6}{64}C_{i,j}+\frac{1}{64}C_{i+1,j}) \nonumber \\
                  &+& \frac{1}{64}C_{i-1,j-1}+\frac{6}{64}C_{i,j-1}+\frac{1}{64}C_{i+1,j-1} \nonumber \\
                  &=& z_{i,j-1}+6z_{i,j}+z_{i,j+1}, 
\end{eqnarray}
where $z_{i,j} \equiv \frac{1}{64}C_{i-1,j}+\frac{6}{64}C_{i,j}+\frac{1}{64}C_{i+1,j}$. }

This suggests to solve a symmetric tridiagonal linear system for $z_{i,j}$:
\begin{equation}
\begin{bmatrix}
    6 & 1 & 0 & 0 & \dots & 0 \\
    1 & 6 & 1 & 0 & \dots & 0 \\ 
    0 & 1 & 6 & 1 & \dots & 0 \\
    \vdots & \vdots & \vdots & \vdots & \ddots & \vdots\\
    0 & 0 & \dots & 1 & 6 & 1 \\
    0 & 0 & \dots & 0 & 1 & 6
\end{bmatrix}
\begin{bmatrix}
z_{i,1} \\
z_{i,2} \\
z_{i,3} \\
\vdots \\
z_{i,N-1} \\
z_{i,N}
\end{bmatrix} = 
\begin{bmatrix}
\sigma_{i,1} \\
\sigma_{i,2} \\
\sigma_{i,3} \\
\vdots \\
\sigma_{i,N-1} \\
\sigma_{i,N}
\end{bmatrix}, 
\label{Eqn:z_system}
\end{equation}
where $i=1,...,M$. One can efficiently solve this matrix using Gaussian elimination. Once the matrix $z_{i,j}$ is solved, we may proceed to solve the a matrix of the same type for $C_{i,j}$:
\hhwang{
\begin{equation}
\dfrac{1}{64}\begin{bmatrix}
    6 & 1 & 0 & 0 & \dots & 0 \\
    1 & 6 & 1 & 0 & \dots & 0 \\ 
    0 & 1 & 6 & 1 & \dots & 0 \\
    \vdots & \vdots & \vdots & \vdots & \ddots & \vdots\\
    0 & 0 & \dots & 1 & 6 & 1 \\
    0 & 0 & \dots & 0 & 1 & 6
\end{bmatrix}
\begin{bmatrix}
C_{1,j} \\
C_{2,j} \\
C_{3,j} \\
\vdots \\
C_{M-1,j} \\
C_{M,j}
\end{bmatrix} = 
\begin{bmatrix}
z_{1,j} \\
z_{2,j} \\
z_{3,j} \\
\vdots \\
z_{M-1,j} \\
z_{M,j}
\end{bmatrix}, 
\label{Eqn:C_system}
\end{equation}
}
where $j=1,...,N$. \hhwang{From Eq.~(\ref{Eqn:sigmaW})  to  Eqs.~(\ref{Eqn:z_system})(\ref{Eqn:C_system}), we have applied the isolated boundary condition $C_{i,j}=0$ when $i<1$ or $j<1$ or $i>M$ or $j>N$.} \hhwang{We note that although the matrix presentation in Eq.~(\ref{Eqn:z_system}) appear to solve for single column $i$, in practice, we may solve for all columns of $z_{i,j}$ independently and simultaneously using Message Passing Interface (MPI) if the density matrix is distributed along one-dimension as the data arrangement needed for using the package FFTW. After the matrix $z_{i,j}$ is solved, a transposing operation is needed before solving the system Eq.~(\ref{Eqn:C_system}). This can be done efficiently using the transposing functions provided by FFTW. } We emphasize that the step of Gaussian elimination is fast, and the associated numerical complexity is linear. This procedure for solving $C_{i,j}$ is systematic and can be easily extended to a higher order method with $k>3$. 

\section{simulations and Results}
In this section, we implement the algorithm detailed in \S~2 for $k=1,2,3$ and compare the results with analytic solutions of the generalized Maclaurin disks, $D_3$ and $D_4$ \citep{wang15}. We will show that the methods implemented for the $B$-spline of order $k=1,2,3$ correspond to the 1st-, 2nd- and 3rd-order of accuracies in terms of the one, two, and infinite norms. 


Now, we have all the ingredients needed for the force calculations, i.e., Eq.~(\ref{eqnF}). \hhwang{To validate the methods, we need analytic density-force pairs that are sufficiently smooth for testing the order of accuracy.} In addition, for exploring the order of accuracy, the mass distribution of disk models has to be finite to be entirely embedded in a computational domain. \hhwang{To our knowledge, the family of generalized Maclaurin disks satisfies both requirements.} 

The surface density of a generalized Maclaurin disk has the form:
\begin{eqnarray}
\sigma_{D_n}(R;\tilde{\alpha})=\begin{cases}
                               \sigma_0\left(1-\dfrac{R^2}{\tilde{\alpha}^2}\right)^{n-1/2}, & \text{~for~}R<\tilde{\alpha} \\
                               0, & \text{~for~}R\geq \tilde{\alpha}, 
                              \end{cases},
\end{eqnarray}
where $R$ is the polar radius, $\sigma_0$ is the disk central density, $\tilde{\alpha}$ represents the disk size and $n=0,1,2,...$ the order of disk. The analytic expressions of gravitational forces for $n=0,1,2$ disks are reported by \citet{Sch09} and generalized to arbitarly positive integers by \citet{wang15, wang16}. The analytic forms of radial forces for $D_3$ and $D_4$ disks in the $z=0$ plane are:
\begin{eqnarray}
F_{R,D_3}(R,0;\eta)=\begin{cases}
                               \dfrac{\pi^2\sigma_0G}{2}\eta^{-5}(-\dfrac{15}{8}\eta^4+\dfrac{45}{16}\eta^2-\dfrac{75}{64}), & \text{~for~}\eta\geq 1 \\
                               \pi\sigma_0G\eta^{-5}\left[\left(-\dfrac{15}{8}\eta^4+\dfrac{45}{16}\eta^2-\dfrac{75}{64}\right)\sin^{-1}(\eta) \right. \\
                               \left. +\left(\dfrac{5}{8}\eta^5-\dfrac{65}{32}\eta^3+\dfrac{75}{64}\eta \right)\sqrt{1-\eta^2}\right] & \text{~for~}\eta \leq 1 
                              \end{cases},
\end{eqnarray}
and
\begin{eqnarray}
F_{R,D_4}(R,0;\eta)=\begin{cases}
                               \dfrac{\pi^2\sigma_0G}{2}\eta^{-7}(-\dfrac{35}{16}\eta^6+\dfrac{315}{64}\eta^4-\dfrac{525}{128}\eta^2+\dfrac{1225}{1024}), & \text{~for~}\eta\geq 1 \\
                               \pi\sigma_0G\eta^{-7}\left[\left(-\dfrac{35}{16}\eta^6+\dfrac{315}{64}\eta^4-\dfrac{525}{128}\eta^2+\dfrac{1225}{1024}\right)\sin^{-1}(\eta) \right. \\
                               \left. +\left(\dfrac{35}{64}\eta^7-\dfrac{1085}{384}\eta^5+\dfrac{5075}{1536}\eta^3-\dfrac{1225}{1024}\eta \right)\sqrt{1-\eta^2}\right] & \text{~for~}\eta \leq 1 
                              \end{cases},
\end{eqnarray}
repectively, and $\eta\equiv \tilde{\alpha}/R$ is a dimensionless variable. These analytic expressions are derived using the recursion formulas Eqs.(53)-(55) in~\citet{wang15}. It will be shown below that the smoothness of a generalized Maclaurin disk is important for tests. In general, the $n$th derivative of $\sigma_{D_n}$ will encounter singularities at the edge of the disk. Thus a $D_n$ disk is not suitable for the test of an algorithm of $n$th-order of accuracy.  That is, to  test an algorithm of $n$th-order accuracy,  one needs at least a disk of $(n+1)$th order. 

Without loss of generality, we set the computational domain $\Omega=[-1,1]\times[-1,1]$,  the parameters for the disks $\sigma_0=G=1$ and  the  disk size $\tilde{\alpha}=0.5$. The $\Omega$ is uniformly subdivided by  $N \times N$  square cells, with $N=32, 64, 128, 256, 512$ being the numbers of zones in both $x$ and  $y$-directions.  The errors compared to the  analytic solutions are measured in terms of the one, two and infinite norms:
\begin{eqnarray}
  L^1 &=& \frac{1}{N^2}\sum_{i=1}^{N}\sum_{j=1}^{N} \left| F^{{\rm num}}_{i,j}-F^{{\rm ana}}_{i,j}\right|, \label{oneNormErr}\\
  L^2 &=& \left(\frac{1}{N^2}\sum_{i=1}^{N}\sum_{j=1}^{N} \left| F^{{\rm num}}_{i,j}-F^{{\rm ana}}_{i,j}\right|^2\right)^{1/2}, \label{twoNormErr}\\
  L^{\infty}&=& {\rm max}(\left| F^{{\rm num}}_{i,j}-F^{{\rm ana}}_{i,j}\right|) \quad\quad \text{for~}i,j=1,...,N, \label{inftyNormErr}
\end{eqnarray}
where $L^1$, $L^2$, $L^{\infty}$ are the one norm, two norm and infinite norm of error, and 
$F^{{\rm num}}_{i,j}$, $F^{{\rm ana}}_{i,j}$ are numerical and exact forces at locations indexed by 
$(i,j)$, respectively. When using $L^1$ and $L^2$, we evaluate the total variation and energy in a global 
sense, while using $L^{\infty}$ we monitor the convergence of maximum errors in a pointwise sense.

\subsection{Results}

Six cases are considered to study the order of accuracy of the algorithm. Tables~\ref{tbl:k1d3} to~\ref{tbl:k3d4} show the numerical results of $(k=1, {D_3} )$, $(k=2, {D_3} )$, $(k=3, {D_3} )$, $(k=1, {D_4} )$, $(k=2, {D_4} )$ and $(k=3, {D_4} )$, respectively. In each case, errors of $x$-forces and radial forces are measured using the one, two and infinity norms, as denoted by $L_{x}^1$, $L_{x}^2$,  $L_{x}^{\infty}$, $L_{R}^1$, $L_{R}^2$,  $L_{R}^{\infty}$, respectively. The upper tables tabulate the numerical values of error measurements with increasing $N$, while the lower tables tabulate the orders of accuracy between different spatial resolutions. 

The results of the case $(k=1,D_3)$ are shown in Table~\ref{tbl:k1d3}. The one norm errors of $x$-forces ($L^1_x$) are listed in the second column in the order of increasing cell numbers, $N$. It is obvious that the numerical accuracy is improved with increasing $N$, i.e., increasing spatial resolution. The degree of improvement for $x$-forces when doubling the spatial resolution is calculated using $O^1_x=\log_2(L^1_{x,N_{p}}/L^1_{x,N_{p+1}})$, where the subscript $p=1,2,3,4$ is used to indicate the comparisons between different cell numbers. For instance, when $(N_p/N_{p+1})=(32/64)$, the $O^1_x = \log_2(\mbox{9.2865e-02/4.4525e-02})=1.06$ measures the degree of improvement, i.e., the order of accuracy of a method using one norm. The order of accuracy depends on how we measure the errors. The most rigorous measurement is the accuracy improvement for the maximum errors, as denoted by $O^{\infty}_x$ in the fourth column. From the second to the fourth columns, we measure the numerical errors using different error norms and list the corresponding order of accuracy for $x$-forces, while from the fifth to the seventh columns, the corresponding measurements are listed for the radial forces. Similar interpretations apply to Tables~\ref{tbl:k2d3} to~\ref{tbl:k3d4}.

\begin{figure}
\epsscale{0.95}
\plotone{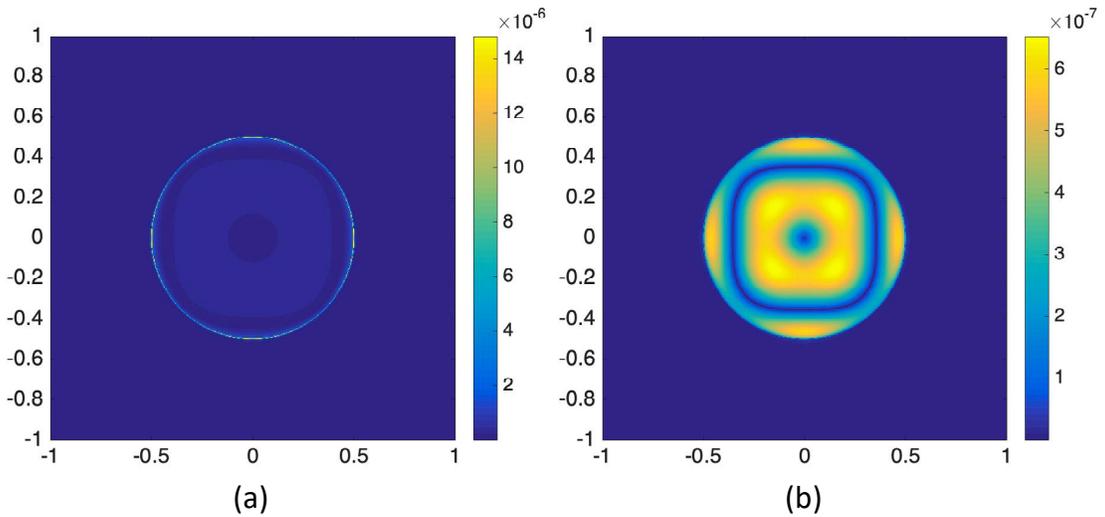}
\caption{The maps of errors of the radial forces for the cases (a) $(k=3, D_3)$ and (b) $(k=3, D_4)$. }
\label{fig:Errmap}
\end{figure}

From Table~\ref{tbl:k1d3} $(k=1,D_3)$, we observe that the order of accuracy converges steadily toward one in terms of one, two and infinite norms. As expected, the result shows that the method using $B$-splines of $k=1$ is of 1st-order accuracy. Similarly, as shown in Table~\ref{tbl:k2d3}$(k=2,D_3)$, the method using $B$-splines of $k=2$ is indeed of 2nd-order accuracy for all error norms. However, in Tabel~\ref{tbl:k3d3}$(k=3,D_3)$, although the method using $B$-splines of $k=3$ converges to 3rd-order accucracy with the one and two norms, $O^{\infty}_x$ and $O^{\infty}_R$ degrade to be less than 2.5. The reason is that the ${D_3}$ disk is not sufficiently smooth at the edge of the disk. This is clearly seen in the Fig.~\ref{fig:Errmap}a, where the maximum errors of the radial forces coincide with the edge of the $D_3$ disk. This also indicates that we need to test the method using $B$-splines of $k=3$ with a smoother disk. We, therefore, adopt the $D_4$ disk for the tests and the results of $k=1,2,3$ are shown in Tables~\ref{tbl:k1d4} to~\ref{tbl:k3d4}, respectively. 

From Tables~\ref{tbl:k1d4} $(k=1,D_4)$ and~\ref{tbl:k2d4}$(k=2,D_4)$, the orders of accuracy converge to one and two, respectively, for all error norms. For $k=1,2$ the $D_4$ disk behaves the same as the $D_3$ disk. However, as shown in Table~\ref{tbl:k3d4} for the case $(k=3, D_4)$, all the error norms are of 3rd-order. If we observe the error distribution for the case $(k=3, D_4)$ shown in Fig.~\ref{fig:Errmap}b, the errors converge more uniformly over the disk. Similar conclusion was also drawn for the $D_2$ disk in \citet{wang15}, where they studied an algorithm of 2nd-order accuracy in polar coordinates.

For a given disk model, one may compare the results between different orders of $B$-splines. We observe that the improvement in numerical accuracy is significant with increasing order of $B$-spline. For example, for the cases with $N=512$, the improvement is more than one order of magnitude between $k=1$ and $k=2$ for both ${D_3}$ and ${D_4}$ disks. The improvement can be comfortably more than two orders of magnitude between $k=2$ and $k=3$ for the ${D_4}$ disk for all error norms. Comparing the gain from accuracy improvement with increasing complexity of the algorithm, it is much more effective to go for an algorithm of higher order. 
\section{Summary and Discussions}
\label{sec:Summary and Discussions}
\subsection{Summary}
In this work, \hhwang{we developed a new algorithm that is fast and of high-order accuracy for calculating self-gravity of infinitesimally thin disks.} The method is fast since the force calculations can be represented by a convolution form, i.e., \hhwang{Eq.~(\ref{eqnF})}. \hhwang{Based on the convolution theorem, a convolution associated with two functions can be evaluated as a pointwise product of their Fourier transforms.} \hhwang{Thus,} the calculation of convolution can be \hhwang{accelerated} by the use of FFT, which is of linear complexity. The high-order accuracy is achieved by approximating the surface density using high-order $B$-splines. The use of $B$-splines has several advantages. First, $B$-splines of high-order are defined recursively. Thus, the mathematical form of basis functions for a high-order method is well-defined. Second, $B$-splines have finite supports. This property directly leads to a linear system with a sparse matrix  described in (\S~2.3), which can be solved efficiently with  linear numerical complexity. Third, the first $k-1$ derivatives of a $B$-spline of order $k$ are continuous. The increasing smoothness with $k$ makes $B$-splines an ideal basis for expanding a smooth function. 

Although the concepts are straightforward, generalizing to high-order of accuracy for practical uses is not obvious. For the calculation of the kernel integral Eq.~(\ref{eqnwB}), we derived the recursion relations Eqs.~(\ref{intF}) and~(\ref{intMoment}). In \S~2.2, detailed calculations are presented for $B$-spline tensors of order $k=1,2,3$. By properly arranging the locations of $B$-spline tensors, we have shown that for $k=1,2$, the calculation of $C_{i,j}=\rho(x_i,y_j)$ is easy, while for $k=3$, a linear system needs to be solved. \hhwang{The linear system is symmetric and sparse \hhwang{\hhwang{with respect to the} diagonal band}. We solve the system using Gaussian elimination, which is also of linear complexity.  \hhwang{All the numerical steps involved in the force calculations are of linear complexity. It concludes that  the overall complexity of the algorithm is therefore linear.} The algorithm can be deployed on a distributed memory system and benefit the parallel computation using MPI.}

To validate the algorithm, we implemented the methods and calculated the self-gravitational forces for the generalized Maclaurin disks $D_3$ and $D_4$. The numerical results are compared with the analytic solutions discussed in \S~3. We adopted the one, two and infinite norms to quantify the order of accuracy. We identified that a $D_3$ disk is not sufficiently smooth for validating a method of 3rd-order. By adopting a $D_4$ disk, we have proved that the method using the $B$-spline tensors of $k=3$ is indeed of 3rd-order accuracy.
 
\subsection{Discussions}
The use of $B$-splines for surface density interpolation has more advantages compared to the Taylor expansion used by \citet{yen12}. First, it reduces the number of performing FFTs. In \citet{yen12}, three times of forward FFTs are needed for both $x-$ and $y$-direction, respectively, and two inverse FFTs are required to get back the $x$ and $y$ forces. That is, for the 2nd-order accuracy, five FFTs  are required before we obtain the self-gravitational forces. For the 3rd-order accuracy using Taylor expansion, eight FFTs need to be performed for every force calculation. However, in this work, two Gaussian eliminations (not required for $k=1,2$), one forward FFT for $C_{i,j}$ and two inverse FFTs for getting the forces are required for {\it all} orders of accuracy. Most of the heavy workload is shifted to the kernel calculations, which needs to be done only once at the beginning of simulations and the results can be reused afterward when needed. 

Second, if one observes the behavior of the order of accuracy for all tables, and compares it with those tables presented by \citet{yen12}, it is obvious that the algorithm using $B$-spline for interpolation is more stable than the algorithm adopting Taylor expansion. Instead of oscillating in the values of order of accuracy as in the work of \citet{yen12}, the method using $B$-splines converges to the expected order smoothly and steadily. In reality, if a disk has discontinuity in the density, the interpolation would introduce the so-called Gibbs phenomenon in the vicinity of discontinuities. However, since the force calculations are related to the surface density through integrations, this side effect can be ignored.

As shown in the Fig.~\ref{fig:Errmap}, adopting an appropriate model that is sufficiently smooth is important for validating a method of a given order of accuracy. For example, for the $B$-splines of $k=2$, a $D_3$ disk is sufficiently smooth for validating a method of 2nd-order accuracy in terms of the infinite norm. However, for $(k=3,D_3)$, the order of accuracy degrades in terms of infinite norm since the edge of the disk is not sufficiently smooth to the third-order. This might not be clearly observed with $L^1$ and $L^2$ since the edge is an one-dimensional structure while these norms evaluate the average values of errors over the whole calculation domain. \hhwang{The squareness of errors seen in Fig.~\ref{fig:Errmap}b is due to the squareness of cells and the squareness of the $B$-spline tensors, which can be also observed in Fig~\ref{fig:Bspline_figure}b. Since our testing cases are axisymmetric and our coordinates are Cartesian, the conclusions drawn are generally applicable to more complicated density distribution.}

\acknowledgements
The authors acknowledge the support of the Theoretical Institute for Advanced Research in Astrophysics (TIARA) based in Academia Sinica Institute of Astronomy and Astrophysics (ASIAA). C.C.Y. thanks the Institute of Astronomy and Astrophysics, Academia Sinica, Taiwan for their constant support. C.C.Y. is supported by Ministry of Science and Technology of Taiwan, under grant MOST-107-2115-M- 030-005-MY2. HHW thanks the support by the grant from the Research Grants Council of Hong Kong: General Research Fund 14308217, and the two supports by the Research Committee Direct Grant for Research from CUHK: 4053229, 4053309. M.-C. Shiue was supported in part by the Grant MOST-106-2115-M-009-011-MY2.

\begin{table}[ht]
\begin{center}
\begin{tabular}{|c|c|}
\hline
range $(x_{\ell}/y_{\ell},x_{r}/y_{r})$ & $(x_{i',0},x_{i',1})$  \\ \hline
$(y_{j',0},y_{j',1})$   & $(1,0,1)$ \\  \hline
\end{tabular}
\caption{Each triplet $(\alpha, \beta, \gamma)$ in the table represents $\gamma\tilde{F}(x_{\ell}, x_r, y_{\ell}, yr, \alpha, \beta)$ defined in Eq.~(\ref{intFtilde}), and $\gamma$ is the associated coefficient. The associated ranges of integration in $x$ and $y$ are indicated in the first row and the first column, respectively. } \label{tbl_Fk1}
\end{center}
\end{table}

\begin{table}[ht]
\begin{center}
\begin{tabular}{|c|c|c|}
\hline
range $(x_{\ell}/y_{\ell},x_{r}/y_{r})$ & $(x_{i',0},x_{i',1})$ & $(x_{i',1},x_{i',2})$ \\ \hline
$(y_{j',0},y_{j',1})$   & $(2,1,1)$, $(1,1,-x_{i',0})$,  &   $(2,1,-1)$, $(2,0,y_{j',0})$ \\ 
                        & $(2,0,-y_{j',0})$, $(1,0,x_{i',0}y_{j',0})$  & $(1,1,x_{i',2})$, $(1,0,-x_{i',2}y_{j',0})$\\ \hline
$(y_{j',1},y_{j',2})$   & $(2,1,-1)$, $(2,0,y_{j',2})$  & $(2,1,1)$, $(2,0,-y_{j',2})$\\ 
                        & $(1,1,x_{i',0})$, $(1,0,-x_{i',0}y_{j',2})$  & $(1,1,-x_{i',2})$, $(1,0,x_{i',2}y_{j',2})$\\ \hline

\end{tabular}
\caption{Each triplet $(\alpha, \beta, \gamma)$ in the table represents $\gamma\tilde{F}(x_{\ell}, x_r, y_{\ell}, yr, \alpha, \beta)$ defined in Eq.~(\ref{intFtilde}), and $\gamma$ is the associated coefficient. The associated ranges of integration in $x$ and $y$ are indicated in the first row and the first column, respectively. } \label{tbl_Fk2}
\end{center}
\end{table}

\begin{sidewaystable}[ht]
\begin{center}
\begin{tabular}{|c|c|c|c|}
\hline
range $(x_{\ell}/y_{\ell},x_{r}/y_{r})$ & $(x_{i',0},x_{i',1})$ & $(x_{i',1},x_{i',2})$ & $(x_{i',2},x_{i',3})$\\ \hline
$(y_{j',0},y_{j',1})$   & $(3,2,1)$, $(3,1,-2y_{0})$, $(3,0,y_{0}^2)$& $(3,2,-2)$, $(3,1,4y_{0})$, $(3,0,-2y_{0}^2)$& $(3,2,1)$, $(3,1,-2y_0)$, $(3,0,y_0^2)$\\ 
                        & $(2,2,-2x_{0})$, $(2,1,4x_{0}y_{0})$, $(2,0,-2x_{0}y_{0}^2)$& $(2,2,A_x)$, $(2,1,-2y_{0}A_x)$, $(2,0,A_xy_0^2)$& $(2,2,-2x_3)$, $(2,1,4x_3y_0)$, $(2,0,-2x_3y_0^2)$\\ 
                        & $(1,2,x_{0}^2)$, $(1,1,-2x_{0}^2y_{0})$, $(1,0,x_{0}^2y_{0}^2)$&$(1,2,-B_x)$,$(1,1,-2B_xy_0)$, $(1,0,-B_xy_0^2)$ & $(1,2,x_3^2)$, $(1,1,-2y_0x_3^2)$, $(1,0,x_3^2y_0^2)$\\ \hline
$(y_{j',1},y_{j',2})$   & $(3,2,-2)$, $(3,1,A_y)$, $(3,0,-B_y)$ & $(3,2,4)$, $(3,1,-2A_y)$, $(3,0,2B_y)$ & $(3,2,-2)$, $(3,1,A_y)$, $(3,0,-B_y)$\\ 
                        & $(2,2,4x_0)$, $(2,1,-2x_0A_y)$, $(2,0,2x_0B_y)$& $(2,2,-2A_x)$, $(2,1,A_xA_y)$, $(2,0,-A_xB_y)$ & $(2,2,4x_3)$, $(2,1,-2A_yx_3)$, $(2,0,2x_3B_y)$\\
                        & $(1,2,-2x_0^2)$, $(1,1,A_yx_0^2)$, $(1,0,-B_yx_0^2)$& $(1,2,2B_x)$, $(1,1,-A_yB_x)$, $(1,0,B_xB_y)$ & $(1,2,-2x_3^2)$, $(1,1,A_yx_3^2)$, $(1,0,-B_yx_3^2)$\\ \hline
$(y_{j',2},y_{j',3})$   & $(3,2,1)$, $(3,1,-2y_3)$, $(3,0,y_3^2)$ & $(3,2,-2)$, $(3,1,4y_3)$, $(3,0,-2y_3^2)$ & $(3,2,1)$, $(3,1,-2y_3)$, $(3,0,y_3^2)$\\ 
                        & $(2,2,-2x_0)$, $(2,1,4x_0y_3)$, $(2,0,-2x_0y_3^2)$ & $(2,2,A_x)$, $(2,1,-2A_xy_3)$, $(2,0,A_xy_3^2)$ & $(2,2,-2x_3)$, $(2,1,4x_3y_3)$, $(2,0,-2x_3y_3^2)$\\
                        & $(1,2,x_0^2)$, $(1,1,-2x_0^2y_3)$, $(1,0,x_0^2y_3^2)$ & $(1,2,-B_x)$, $(1,1,2B_xy_3)$, $(1,0,-B_xy_3^2)$ & $(1,2,x_3^2)$, $(1,1,-2x_3^2y_3)$, $(1,0,x_3^2y_3^2)$\\ \hline                       

\end{tabular}
\caption{Each triplet $(\alpha, \beta, \gamma)$ in the table represents $\gamma\tilde{F}(x_{\ell}, x_r, y_{\ell}, yr, \alpha, \beta)$ defined in Eq.~(\ref{intFtilde}), and $\gamma$ is the associated coefficient. The associated ranges of integration in $x$ and $y$ are indicated in the first row and the first column, respectively. To avoid the cluttering symbols, we have suppressed the subscripts $i'$ and $j'$  in the triplets. In this table, $A_x=x_0+x_1+x_2+x_3$, $B_x=x_0x_2+x_1x_3$, $A_y=y_0+y_1+y_2+y_3$, $B_y=y_0y_2+y_1y_3$.} \label{tbl_Fk3}
\end{center}
\end{sidewaystable}

\begin{table}[ht]
\begin{center}
\begin{tabular}{|c|c|c|c|c|c|c|}
\hline
$N$ & $L_{x}^1$ & $L_{x}^2$ & $L_{x}^{\infty}$ & $L_{R}^1$ & $L_{R}^2$ & $L_{R}^{\infty}$ \\ \hline
32   & 9.2865e-02  &1.3093e-01   &2.9581e-01   & 1.4516e-01 & 1.8516e-01 & 3.0346e-01  \\ \hline
64   & 4.4525e-02  &6.3186e-02  &1.4563e-01     &6.9853e-02& 8.9359e-02 &1.4666e-01\\ \hline
128  & 2.1765e-02  &3.0984e-02  &7.1745e-02     &3.4178e-02&4.3819e-02&7.1872e-02\\ \hline
256 & 1.0757e-02  &1.5336e-02   &3.5542e-02     &1.6895e-02&2.1688e-02&3.5558e-02\\ \hline
512 & 5.3472e-03  &7.6281e-03   &1.7681e-02     & 8.3991e-03&1.0788e-02&1.7683e-02\\ \hline
\hline
$N_p/N_{p+1}$ &  $O_{x}^1$ & $O_{x}^2$ & $O_{x}^{\infty}$ & $O_{R}^1$ & $O_{R}^2$ & $O_{R}^{\infty}$ \\ \hline
32/64           &  1.06    & 1.05     &  1.02      & 1.06     & 1.05     & 1.05 \\ \hline
64/128         &   1.03    & 1.03     &   1.02     & 1.03     & 1.03     & 1.03 \\ \hline
128/256       &  1.02     &  1.01    &   1.01     &  1.02    &   1.01   &1.02 \\ \hline
256/512       &  1.01     &   1.01   &    1.01    &  1.01    &  1.01    &1.01 \\ \hline
\end{tabular}
\caption{This table tabulates the errors and orders of accuracy of the $x$ and radial forces for the case $(k=1, {D_3})$}  \label{tbl:k1d3}
\end{center}
\end{table}

\begin{table}[ht]
\begin{center}
\begin{tabular}{|c|c|c|c|c|c|c|}
\hline
$N$ & $L_{x}^1$ & $L_{x}^2$ & $L_{x}^{\infty}$ & $L_{R}^1$ & $L_{R}^2$ & $L_{R}^{\infty}$ \\ \hline
32   & 1.7712e-02  & 2.0437e-02   &4.5876e-02   & 2.7583e-02 & 2.8873e-02 & 4.6383e-02 \\ \hline
64   & 4.5916e-03  &5.3308e-03  &1.2064e-02     & 7.1866e-03 &7.5370e-03 & 1.2085e-02\\ \hline
128  & 1.1689e-03  &1.3609e-03  &3.0684e-03     &1.8330e-03 &1.9244e-03 &3.0698e-03\\ \hline
256 & 2.9472e-04  &3.4359e-04   &7.7234e-04    & 4.6251e-04& 4.8589e-04&7.7258e-04 \\ \hline
512 & 7.3986e-05  &8.6311e-05   &1.9378e-04     & 1.1614e-04 &1.2206e-04 &1.9379e-04 \\ \hline
\hline
$N_p/N_{p+1}$ &  $O_{x}^1$ & $O_{x}^2$ & $O_{x}^{\infty}$ & $O_{R}^1$ & $O_{R}^2$ & $O_{R}^{\infty}$ \\ \hline
32/64           &  1.95    & 1.94     &  1.93     & 1.94     & 1.94    & 1.94 \\ \hline
64/128         &   1.97   &   1.97   &    1.98    & 1.97    &   1.97   &  1.98\\ \hline
128/256       &  1.99    &   1.99  &    1.99    &  1.99   &  1.99   & 1.99\\ \hline
256/512       &  1.99   &   1.99  &  1.99   & 1.99   & 1.99  &  2.00 \\ \hline
\end{tabular}
\caption{This table tabulates the errors and orders of accuracy of the $x$ and radial forces for the case $(k=2, {D_3})$} \label{tbl:k2d3}
\end{center}
\end{table}

\begin{table}[ht]
\begin{center}
\begin{tabular}{|c|c|c|c|c|c|c|}
\hline
$N$ & $L_{x}^1$ & $L_{x}^2$ & $L_{x}^{\infty}$ & $L_{R}^1$ & $L_{R}^2$ & $L_{R}^{\infty}$ \\ \hline
32   & 2.0387e-03  & 3.0324e-03  & 1.2152e-02 & 3.3411e-03 & 4.1530e-03 & 1.1409e-02 \\ \hline
64   & 1.9115e-04  & 3.7919e-04 & 2.4028e-03   & 3.1242e-04  &5.2093e-04 &2.3122e-03 \\ \hline
128  & 1.9873e-05  & 4.8220e-05 & 4.6035e-04  & 3.1787e-05 &6.6308e-05 &4.5209e-04\\ \hline
256 &  2.1659e-06 & 6.0452e-06  & 8.3088e-05   &3.4478e-06 & 8.3172e-06& 8.2252e-05\\ \hline
512 & 2.4457e-07  & 7.5529e-07  & 1.4890e-05    & 3.8775e-07 &1.0381e-06 & 1.4826e-05\\ \hline
\hline
$N_p/N_{p+1}$ &  $O_{x}^1$ & $O_{x}^2$ & $O_{x}^{\infty}$ & $O_{R}^1$ & $O_{R}^2$ & $O_{R}^{\infty}$ \\ \hline
32/64           &  3.41    & 3.00     & 2.34      & 3.42     & 3.00    &  2.30\\ \hline
64/128         &   3.27   &  2.98    &  2.38      & 3.30    &  2.97    & 2.35 \\ \hline
128/256       &  3.20    &  3.00   &   2.47     & 3.20    & 3.00    & 2.46 \\ \hline
256/512       & 3.15    & 3.00    & 2.48    &  3.15  & 3.00  & 2.47 \\ \hline
\end{tabular}
\caption{This table tabulates the errors and orders of accuracy of the $x$ and radial forces for the case $(k=3, {D_3})$} \label{tbl:k3d3}
\end{center}
\end{table}

\begin{table}[ht]
\begin{center}
\begin{tabular}{|c|c|c|c|c|c|c|}
\hline
$N$ & $L_{x}^1$ & $L_{x}^2$ & $L_{x}^{\infty}$ & $L_{R}^1$ & $L_{R}^2$ & $L_{R}^{\infty}$ \\ \hline
32   & 8.1599e-02  & 1.2740e-01   & 3.3661e-01 &1.2736e-01  & 1.8017e-01 & 3.4076e-01 \\ \hline
64   & 3.8941e-02  & 6.1282e-02 &  1.6416e-01  & 6.1065e-02 & 8.6666e-02&1.6471e-01 \\ \hline
128  & 1.9026e-02  & 2.9995e-02 & 8.0277e-02 & 2.9873e-02&4.2419e-02 &8.0450e-02\\ \hline
256 & 9.4065e-03 & 1.4831e-02  & 3.9722e-02  &1.4774e-02 & 2.0974e-02&3.9740e-02 \\ \hline
512 &  4.6772e-03 & 7.3733e-03  & 1.9744e-02  &7.3467e-03  &1.0427e-02 &1.9746e-02 \\ \hline
\hline
$N_p/N_{p+1}$ &  $O_{x}^1$ & $O_{x}^2$ & $O_{x}^{\infty}$ & $O_{R}^1$ & $O_{R}^2$ & $O_{R}^{\infty}$ \\ \hline
32/64           & 1.07     & 1.06     &  1.04     & 1.06     & 1.06    & 1.05 \\ \hline
64/128         &  1.03    &  1.03    &   1.03     & 1.03    &  1.03    & 1.03 \\ \hline
128/256       &  1.02    &  1.02   &  1.02      & 1.02    & 1.02    & 1.02\\ \hline
256/512       &  1.01   &  1.01   & 1.01    & 1.01   & 1.01  &1.01 \\ \hline
\end{tabular}
\caption{This table tabulates the errors and orders of accuracy of the $x$ and radial forces for the case $(k=1, {D_4})$}  \label{tbl:k1d4}
\end{center}
\end{table}

\begin{table}[ht]
\begin{center}
\begin{tabular}{|c|c|c|c|c|c|c|}
\hline
$N$ & $L_{x}^1$ & $L_{x}^2$ & $L_{x}^{\infty}$ & $L_{R}^1$ & $L_{R}^2$ & $L_{R}^{\infty}$ \\ \hline
32   & 1.5794e-02  & 2.1363e-02   & 5.8539e-02 & 2.4639e-02 & 3.0197e-02 & 5.9092e-02 \\ \hline
64   &  4.1209e-03 & 5.5802e-03 &  1.5555e-02  & 6.4603e-03 &7.8906e-03 & 1.5600e-02 \\ \hline
128  & 1.0494e-03  & 1.4225e-03 &  3.9659e-03  & 1.6471e-03 & 2.0116e-03  & 3.9692e-03\\ \hline
256 & 2.6472e-04 & 3.5886e-04  &  1.0002e-03  & 4.1563e-04 &5.0750e-04 & 1.0004e-03\\ \hline
512 &  6.6463e-05 & 9.0107e-05  &  2.5103e-04   & 1.0436e-04 & 1.2743e-04& 2.5105e-04\\ \hline
\hline
$N_p/N_{p+1}$ &  $O_{x}^1$ & $O_{x}^2$ & $O_{x}^{\infty}$ & $O_{R}^1$ & $O_{R}^2$ & $O_{R}^{\infty}$ \\ \hline
32/64           &  1.94    &  1.94    & 1.91    & 1.93     &  1.94   & 1.92 \\ \hline
64/128         &   1.97   &   1.97   &   1.97     & 1.97    &  1.97    & 1.97 \\ \hline
128/256       &  1.99    &  1.99   &   1.99     & 1.99    &  1.99   & 1.99\\ \hline
256/512       &  1.99   &  1.99   &  1.99   & 1.99   & 1.99  & 1.99\\ \hline
\end{tabular}
\caption{This table tabulates the errors and orders of accuracy of the $x$ and radial forces for the case $(k=2, {D_4})$} \label{tbl:k2d4}
\end{center}
\end{table}

\begin{table}[ht]
\begin{center}
\begin{tabular}{|c|c|c|c|c|c|c|}
\hline
$N$ & $L_{x}^1$ & $L_{x}^2$ & $L_{x}^{\infty}$ & $L_{R}^1$ & $L_{R}^2$ & $L_{R}^{\infty}$ \\ \hline
32   & 1.2650e-03  & 1.6606e-03  & 4.9032e-03  & 1.9866e-03 & 2.2916e-03 &  4.8862e-03\\ \hline
64   & 1.2213e-04  & 1.6468e-04 & 4.4374e-04   & 1.8915e-04 & 2.2746e-04 & 4.3164e-04\\ \hline
128  & 1.3308e-05  & 1.8216e-05 &  4.4007e-05   & 2.0463e-05 & 2.5149e-05 & 4.4150e-05\\ \hline
256 &  1.5530e-06 & 2.1442e-06  & 4.9395e-06   & 2.3780e-06 &2.9586e-06 & 5.3249e-06\\ \hline
512 &  1.8790e-07 &  2.6036e-07 & 5.9366e-07    & 2.8684e-07 &3.5910e-07 &6.5341e-07 \\ \hline
\hline
$N_p/N_{p+1}$ & $O_{x}^1$ & $O_{x}^2$ & $O_{x}^{\infty}$ & $O_{R}^1$ & $O_{R}^2$ & $O_{R}^{\infty}$ \\ \hline
32/64           &  3.37    &  3.33    &  3.47     &  3.39    &  3.33   &  3.50 \\ \hline
64/128         &  3.20    &  3.18    &   3.33     & 3.21    &  3.18    &  3.29 \\ \hline
128/256       &  3.10    &  3.09   &   3.16     &  3.11   & 3.09    & 3.05 \\ \hline
256/512       &  3.05   &  3.04   &  3.06   & 3.05   & 3.04  & 3.03\\ \hline
\end{tabular}
\caption{This table tabulates the errors and orders of accuracy of the $x$ and radial forces for the case $(k=3, {D_4})$} \label{tbl:k3d4}
\end{center}
\end{table}


\bibliography{BS_gravity}

\appendix
\section{Formula}
For the integral $\displaystyle F(x,y,\alpha,\beta)=\int\!\!\!\int \frac{x^\alpha y^\beta}{(x^2+y^2)^{3/2}}\,dxdy$ with $\alpha$ and $\beta$, some of its antiderivatives are listed below.
\begin{table}[ht]
\begin{center}
\begin{tabular}{ccl}
\hline \hline\\
$\alpha$ & $\beta$ & $F(x,y,\alpha,\beta)$
\\ \hline \hline\\
0   & 0   & $\displaystyle -\frac{\sqrt{x^2+y^2}}{xy}+C     $
\\ \hline\\
1   & 0   & $\displaystyle -\ln (y+\sqrt{x^2+y^2})+C        $ \\
0   & 1   & $F(y,x,1,0)$
\\  \hline\\
2   & 0   & $\displaystyle y\ln (x+\sqrt{x^2+y^2})+C        $\\
1   & 1   & $\displaystyle -\sqrt{x^2+y^2}+C                $\\
0   & 2   & $F(y,x,2,0)$
\\ \hline\\
3   & 0   & $\displaystyle y\sqrt{x^2+y^2}+C                $\\
2   & 1   & $\displaystyle -\frac{1}{2}x\sqrt{x^2+y^2}
+\frac{1}{2}y^2\ln(x+\sqrt{x^2+y^2})+C                      $\\
1   & 2   & $F(y,x,2,1)$\\
0   & 3   & $F(y,x,3,0)$
\\ \hline\\
4   & 0   & $\frac{1}{2}xy\sqrt{x^2+y^2}-\frac{1}{2}y^3\ln(x+\sqrt{x^2+y^2})+C$ \\
3   & 1   & $-\frac{1}{3}\sqrt{x^2+y^2}(x^2-2y^2)+C$  \\
2   & 2   & $-\frac{1}{3}xy\sqrt{x^2+y^2}+\frac{1}{3}y^3\ln(x+\sqrt{x^2+y^2})+\frac{1}{3}x^3\ln(y+\sqrt{x^2+y^2})+C$ \\
1   & 3   & $F(y,x,3,1)$\\
0   & 4   & $F(y,x,4,0)$
\\ \hline
\end{tabular}
\caption{This table shows the kernel integrals for $\alpha+\beta=0,1,2,3,4$.}
\end{center}
\end{table}

\section{Derivation of Equations~(15)(18)}
Before deriving Eq.~(15), we shall derive Eq.~(18) first. 
\begin{eqnarray}
\int \frac{x^{\alpha}}{(x^2+y^2)^{3/2}} dx&=& -\frac{x^{\alpha-1}}{(x^2+y^2)^{1/2}}+(\alpha-1)\int \frac{x^{\alpha-2}(x^2+y^2)}{(x^2+y^2)^{3/2}}dx \\
 &=& -\frac{x^{\alpha-1}}{(x^2+y^2)^{1/2}}+(\alpha-1)y^2\int\frac{x^{\alpha-2}}{(x^2+y^2)^{3/2}}+(\alpha-1)\int \frac{x^{\alpha}}{(x^2+y^2)^{3/2}} dx,
\end{eqnarray}

where Eq.~(B1) is obtained using the integration by part and the integral in the third term of Eq.~(B2) is the same as the original integral. From Eq.~(B2), we \hhwang{obtain} a recursion relation, Eq.~(18):
\begin{eqnarray}
\int \frac{x^{\alpha}}{(x^2+y^2)^{3/2}} dx = \frac{1}{\alpha-2}\frac{x^{\alpha-1}}{(x^2+y^2)^{1/2}}-\frac{\alpha-1}{\alpha-2}y^2\int \frac{x^{\alpha-2}}{(x^2+y^2)^{3/2}} dx. 
\end{eqnarray}

Now, we are in a position to derive Eq.~(15). To begin with,
\begin{eqnarray}
\int\int\frac{x^{\alpha}y^{\beta-2}}{(x^2+y^2)^{3/2}}dxdy&=&-\int\frac{x^{\alpha-1}y^{\beta-2}(x^2+y^2)}{(x^2+y^2)^{3/2}}dy+(\alpha-1)\int\int\frac{x^{\alpha-2}y^{\beta-2}(x^2+y^2)}{(x^2+y^2)^{3/2}}dxdy \\
&=&-x^{\alpha+1}\int\frac{y^{\beta-2}}{(x^2+y^2)^{3/2}}dy-x^{\alpha-1}\int\frac{y^{\beta}}{(x^2+y^2)^{3/2}}dy \nonumber \\
&&+(\alpha-1)\int\int\frac{x^{\alpha}y^{\beta-2}}{(x^2+y^2)^{3/2}}dxdy+(\alpha-1)\int\int\frac{x^{\alpha-2}y^{\beta}}{(x^2+y^2)^{3/2}}dxdy,
\end{eqnarray}
where Eq.~(B4) is obtained by multiplying Eq.~(B1) with $y^{{\beta-2}}$
and integrating the resulting in the third term in Eq.~(B5) is the same as 
the original one and the integral in the fourth term in Eq.~(B5) is symmetric 
to the original one, it can be inferred from Eq.~(B5) that
\begin{eqnarray}
\int\int\frac{x^{\alpha}y^{\beta-2}}{(x^2+y^2)^{3/2}}dxdy&=& F(x,y, \alpha, \beta-2) \nonumber \\
&=&\frac{x^{\alpha+1}}{\alpha-2}\int\frac{y^{\beta-2}}{(x^2+y^2)^{3/2}}dy+\frac{x^{\alpha-1}}{\alpha-2}\int\frac{y^{\beta}}{(x^2+y^2)^{3/2}}dy \nonumber \\
&&-\frac{\alpha-1}{\alpha-2}F(x,y,\alpha-2,\beta). 
\end{eqnarray}
By the same token, we obtain a relation for the symmetric expression:
\begin{eqnarray}
\int\int\frac{x^{\alpha-2}y^{\beta}}{(x^2+y^2)^{3/2}}dxdy&=& F(x,y, \alpha-2, \beta) \nonumber \\
&=&\frac{y^{\beta+1}}{\beta-2}\int\frac{x^{\alpha-2}}{(x^2+y^2)^{3/2}}dx+\frac{y^{\beta-1}}{\beta-2}\int\frac{x^{\alpha}}{(x^2+y^2)^{3/2}}dx \nonumber \\
&&-\frac{\beta-1}{\beta-2}F(x,y,\alpha,\beta-2). 
\end{eqnarray}
Substituting Eq.~(B6) into Eq.~(B7), we have:
\begin{eqnarray}
F(x,y,\alpha,\beta)&=&-\frac{\alpha}{\alpha+\beta-1}y^{\beta-1}\int\frac{x^{\alpha+2}}{(x^2+y^2)^{3/2}}dx-\frac{\alpha}{\alpha+\beta-1}y^{\beta+1}\int\frac{x^{\alpha}}{(x^2+y^2)^{3/2}}dx \nonumber \\
&&+\frac{\beta-1}{\alpha+\beta-1}x^{\alpha+1}\int\frac{y^{\beta}}{(x^2+y^2)^{3/2}}dy+\frac{\beta-1}{\alpha+\beta-1}x^{\alpha+3}\int\frac{y^{\beta-2}}{(x^2+y^2)^{3/2}}dy.
\end{eqnarray}
Substituting Eq.~(B7) into Eq.~(B6), we have another expression:
\begin{eqnarray}
F(x,y,\alpha,\beta)&=&-\frac{\beta}{\alpha+\beta-1}x^{\alpha-1}\int\frac{y^{\beta+2}}{(x^2+y^2)^{3/2}}dy-\frac{\beta}{\alpha+\beta-1}x^{\alpha+1}\int\frac{y^{\beta}}{(x^2+y^2)^{3/2}}dy \nonumber \\
&&+\frac{\alpha-1}{\alpha+\beta-1}y^{\beta+1}\int\frac{x^{\alpha}}{(x^2+y^2)^{3/2}}dx+\frac{\alpha-1}{\alpha+\beta-1}y^{\beta+3}\int\frac{x^{\alpha-2}}{(x^2+y^2)^{3/2}}dx.
\end{eqnarray} 
An expression that reflects the symmetry in $x$ and $y$ is obtained by $F(x,y,\alpha,\beta)=(\mbox{B8+B9})/2$: 

\hhwang{
\begin{eqnarray}
F(x,y,\alpha,\beta) &=& \frac{y^{\beta+1}}{\alpha+\beta-1}\int \frac{x^{\alpha}}{(x^2+y^2)^{3/2}}dx + \frac{x^{\alpha+1}}{\alpha+\beta-1}\int \frac{y^{\beta}}{(x^2+y^2)^{3/2}}dy\nonumber \\
&=&  \frac{y^{\beta+1}}{\alpha+\beta-1} G(x,y,\alpha)+
\frac{x^{\alpha+1}}{\alpha+\beta-1}G(y,x,\beta)
\end{eqnarray}
}

Since the close form of the integral Eq.~(B3) exists for $\alpha=0,1,2$, the evaluation of Eq.~(B10) is straightforward. \hhwang{As a reference, the analytic expressions for $\alpha=0,1,2$ are listed below:
\begin{eqnarray}
\int \frac{1}{(x^2+y^2)^{3/2}}dx = \frac{x}{y^2\sqrt{x^2+y^2}} +C, \\
\int \frac{x}{(x^2+y^2)^{3/2}}dx = -\frac{1}{\sqrt{x^2+y^2}} +C, \\
\int \frac{x^2}{(x^2+y^2)^{3/2}}dx = -\frac{x}{\sqrt{x^2+y^2}}+\ln |x+\sqrt{x^2+y^2}| +C,
\end{eqnarray}
where $C$ is an integration constant. For example, to evaluate $F(x,y,2,1)$ from Eq. (B10):
\begin{eqnarray}
F(x,y,2,1) &=& \frac{y^2}{2}\int \frac{x^2}{(x^2+y^2)^{3/2}}dx + \frac{x^3}{2}\int \frac{y}{(x^2+y^2)^{3/2}}dy \\
&=& \frac{y^2}{2}\left(-\frac{x}{\sqrt{x^2+y^2}}+\ln|x+\sqrt{x^2+y^2}| \right) + \frac{x^3}{2}\left( - \frac{1}{\sqrt{x^2+y^2}}\right) +C \\
&=& -\frac{x}{2}\sqrt{x^2+y^2}+\frac{y^2}{2}\ln |x+\sqrt{x^2+y^2}| + C.
\end{eqnarray}
From Eq. (B14) to Eq. (B15), we have used Eqs. (B12) and (B13). To illustrate the use of Eq. (B3), we evaluate $F(x,y,4,0)$ from (B10) as follows:
\begin{eqnarray}
F(x,y,4,0) &=& \frac{y}{3}\int \frac{x^4}{(x^2+y^2)^{3/2}}dx + \frac{x^5}{3}\int \frac{1}{(x^2+y^2)^{3/2}}dy \\
 &=& \frac{y}{3}\left(\frac{1}{2}\frac{x^3}{\sqrt{x^2+y^2}} - \frac{3}{2} y^2 \int \frac{x^2}{(x^2+y^2)^{3/2}}dx \right) + \frac{x^5}{3}\frac{y}{x^2\sqrt{x^2+y^2}}+C\\
 &=& \frac{1}{2}xy\sqrt{x^2+y^2}-\frac{y^3}{2}\ln |x+\sqrt{x^2+y^2}|+C.
\end{eqnarray}
From Eq. (B17) to Eq. (B18), we have applied Eq. (B3) to evaluate the first integral and Eq. (B11) to the second. From Eq. (B18) to (B19), we have applied (B13) to the integral. 

For a general $F(x, y, \alpha, \beta)$, the evaluation procedure is the following:
\begin{enumerate}
\item Apply Eq. (B10). 
\item If the exponent of the integrand numerator is larger than 2, apply Eq. (B3)  to reduce the exponent by 2. 
\item If the exponent of the integrand numerator is equal or less than 2, apply Eqs. (B11) to (B13) directly; go back to step 2, otherwise.
\end{enumerate}

}

\end{document}